\documentclass{PoS}
\usepackage{dcolumn}
\usepackage{bm}
\usepackage{color}
\usepackage{amssymb}
\usepackage{amsmath}
\usepackage{graphicx}
\usepackage{pstricks}
\allowdisplaybreaks
\usepackage{float}
\title{$\chi$EFT studies of few-nucleon systems: a status report}

\ShortTitle{$\chi$EFT studies of light nuclei}

\author{\speaker{R. Schiavilla}\\
        Theory Center, Jefferson Lab, Newport News, VA23606\\
         Department of Physics, Old Dominion University, Norfolk, VA 23529\\
        E-mail: \email{schiavil@jlab.org}}

\abstract{A status report on $\chi$EFT studies of few-nucleon
electroweak structure and dynamics is provided, including
electromagnetic form factors of few-nucleon systems,
the $pp$ weak fusion and muon weak captures on deuteron and
$^3$He, and a number of parity-violating processes induced by
hadronic weak interactions.
}

\FullConference{The 8th International Workshop on Chiral Dynamics, CD2015 ***\\
		29 June 2015 - 03 July 2015\\
		Pisa,Italy}
\begin{document}
\section{Introduction}
\label{sec:intro}
In this talk we review some recent applications of
nuclear chiral effective field theory ($\chi$EFT) to
electroweak  properties of few-nucleon
systems.  We review briefly the derivation of
nuclear potentials and electroweak currents from
chiral Lagrangians.  The currents contain low-energy
constants (LEC's) that need to be fixed.  We
review the various strategies we have been exploring to this end.

The main part of the talk discusses applications to
electromagnetic structure,
weak transitions, and hadronic
weak interactions.  We conclude with a brief
summary and outlook into the next stage from
the perspective of our group.

\section{Nuclear $\chi$EFT}
\label{sec:chieft}

Chiral effective field theory is a low-energy approximation
of quantum chromodynamics (QCD).  The (approximate) chiral
symmetry satisfied by QCD requires that pions couple to nucleons and other
pions by powers of their momenta $Q$, and the Lagrangian
describing the interactions of these particles can be
expanded in powers of $Q/\Lambda_\chi$, where $\Lambda_\chi$
is a ``hard'' scale ($\Lambda_\chi \sim 1$ GeV).  By now,
the construction of these Lagrangians has been codified in a number
of classic papers~\cite{Gasser84,Fettes00} (several of their authors
are attending this conference).

The application to nuclear physics requires going beyond perturbation
theory in order to deal with bound states.  Weinberg suggested to
construct the two-nucleon potential by only considering irreducible
contributions to the scattering amplitude~\cite{Weinberg90}. Reducible contributions
are generated by solutions of the Lippmann-Schwinger (LS) equation.  Applications
of this framework to nuclear potentials~\cite{Ord95,Epelbaum98} and electroweak currents~\cite{Park93,Park96}
soon followed.

Our formalism, as well as Weinberg's, for constructing nuclear potentials
and currents is based on time-ordered perturbation theory (TOPT)~\cite{Pastore08}.  Terms
in this expansion are represented by diagrams, each characterized by
a  number of vertices, energy denominators, and loops.  These elements
scale with a certain power of $Q$, the low-momentum scale.  A diagram
will generally have energy denominators involving only nucleon kinetic
energies, which scale as $Q^2$, and energy denominators involving
in addition pion energies, which scale as $Q$.  The latter are expanded as 
\[
\frac{1}{E_i-E_I-\omega_\pi} = -\frac{1}{\omega_\pi}\left[ 1 + \frac{E_i-E_I}{\omega_\pi} +
\frac{(E_i-E_I)^2}{\omega^2_\pi} + \dots\right]  \ ,
\]
and the leading order term $-1/\omega_\pi$ is the static correction,
while the remaining terms represent non-static corrections of increasing
order $Q^1$, $Q^2$, and so on.  This permits the expansion of the amplitude
($T$-matrix) in a power series. 

The two-nucleon potential is constructed by requiring that its iteration in the LS equation,
\[
v+v\, G_0\, v+v\, G_0 \, v\, G_0 \, v +\dots \ ,  
\]
where $G_0=1/(E_i-E_I+i\, \eta)$ and 
$v=v^{(0)}+v^{(1)}+v^{(2)}+\dots$ with $v^{(n)}\sim Q^n$,
matches the field-theory amplitude order by order in the power counting~\cite{Pastore11}.
This matching leads to the following relations:
\begin{eqnarray}
\hspace{-1cm}v^{(0)} &=& T^{(0)} \ , \nonumber \\
\hspace{-1cm}v^{(1)} &=& T^{(1)}-\left[ v^{(0)}\, G_0\, v^{(0)}\right] \ , \nonumber\\
\hspace{-1cm}v^{(2)} &=&  T^{(2)}-\left[ v^{(0)}\, G_0\, v^{(0)}\, G_0\, v^{(0)}\right] 
 -\left[ v^{(1)}\, G_0 \, v^{(0)}
+v^{(0)}\, G_0\, v^{(1)}\right] \nonumber 
\end{eqnarray}
and so on, where $v^{(m)}\, G_0 \, v^{(n)} \sim Q^{m+n+1}$.  There are in general
partial cancellations between the field theory amplitude and the iterations of
lower order potentials.  The ``left-overs", which we denote as recoil corrections,
are ignored in Weinberg's scheme, and this is the main difference between
his and our approach.

Inclusion of electroweak interactions, which are treated in first order
in the perturbative expansion, is a straightforward extension of the method
just discussed~\cite{Pastore11,Pastore09,Piarulli13,Baroni15}.  In
the electromagnetic (EM) case, the field theory amplitude $T_\gamma$
has the expansion
$T_\gamma=T_\gamma^{(-3)}+T_\gamma^{(-2)}+T_\gamma^{(-1)} +\dots\,\,$~\cite{Park96}.
We define an EM potential $v_\gamma=A^0\, \rho-{\bf A}\cdot {\bf j}$,
where $A^\mu=(A^0,{\bf A})$ is the external EM field and
$\rho$ and ${\bf j}$ are the nuclear charge and current operators,
and do the matching order by order to obtain
\begin{eqnarray}
v_\gamma^{(-3)}&=& T_\gamma^{(-3)} \ , \nonumber \\
v_\gamma^{(-2)}&=& T_\gamma^{(-2)}-\left[ v_\gamma^{(-3)}\, G_0\, v^{(0)}+
 v^{(0)}\, G_0\, v_\gamma^{(-3)} \right] \ , \nonumber \\
 v_\gamma^{(-1)}&=&T_\gamma^{(-1)}
-\left[ v_\gamma^{(-3)}\, G_0\, v^{(0)}\, G_0\, v^{(0)}
+{\rm permutations} \right] \nonumber \\
&&\qquad\,\,\,\, -
\left[v_\gamma^{(-2)}\, G_0\, v^{(0)}+v^{(0)}\, G_0\, v_\gamma^{(-2)}\right] \nonumber \ ,
\end{eqnarray}
\begin{figure}[bth]
\includegraphics[width=6.75cm]{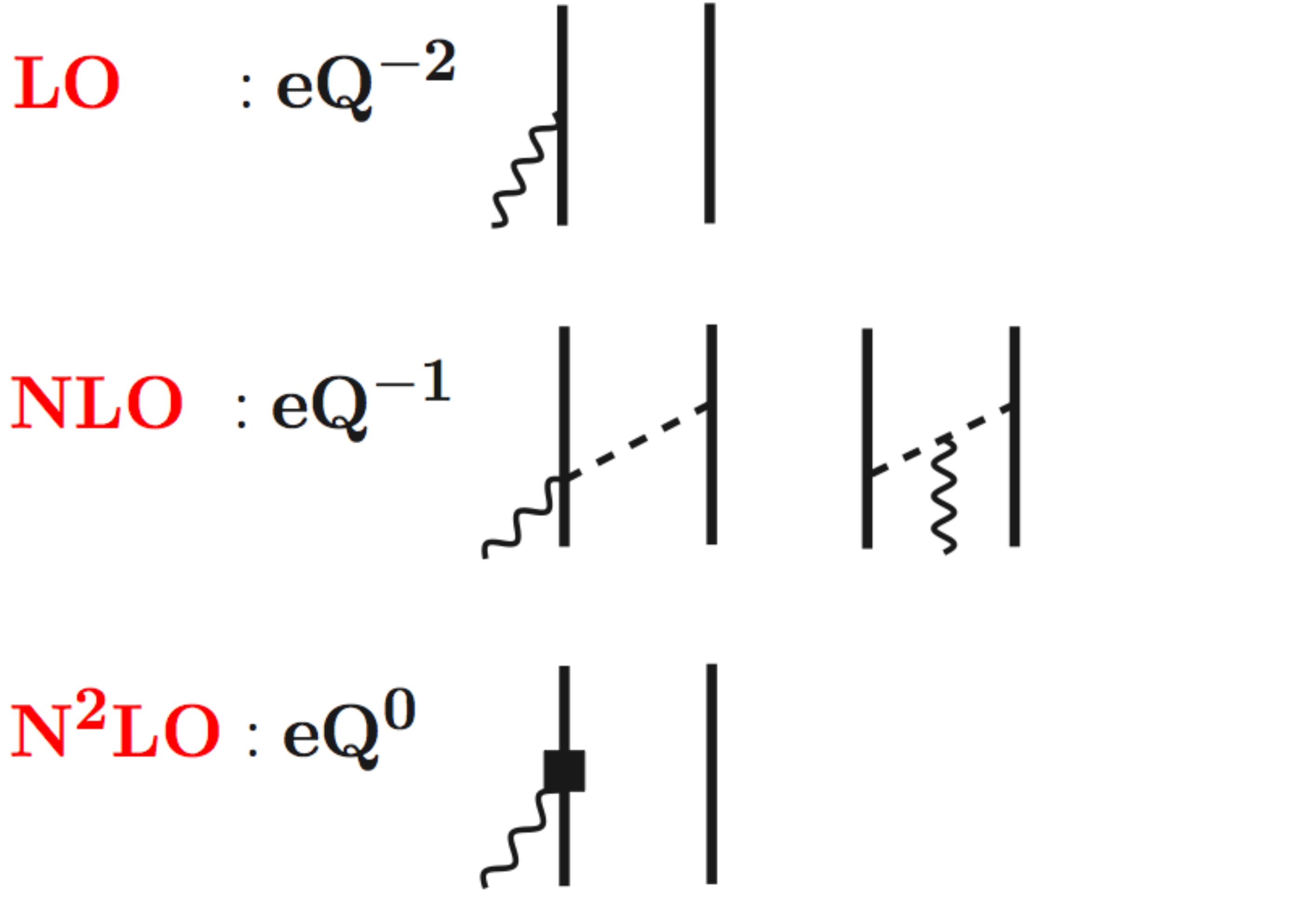}\\
\vspace{0.2cm}
\hspace{-0.20cm}
\includegraphics[width=14.25cm]{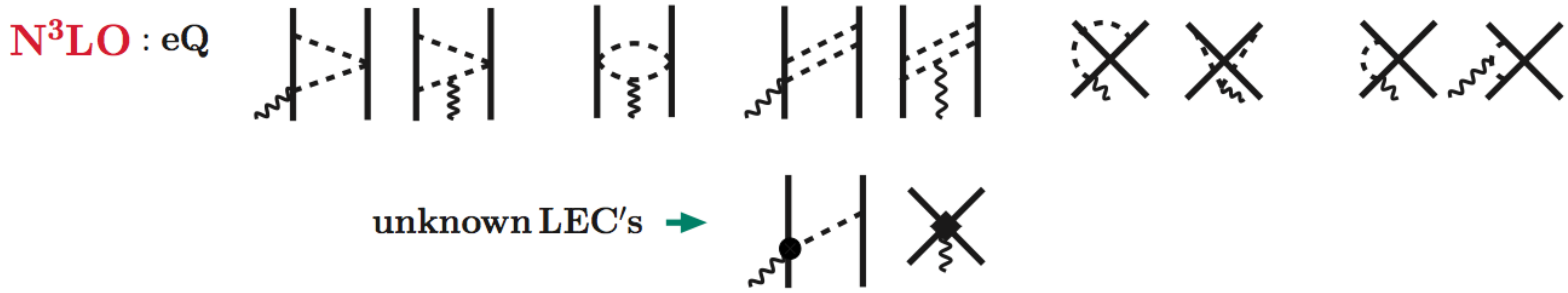}
\caption{Contributions to the nuclear EM current ${\bf j}^{(n)}$.  Nucleons, pions, and photons
are represented by solid, dashed, and wavy lines, respectively.}
\label{fig:f1}
\end{figure}
and so on.  These relations determine $\rho^{(n)}$, the leading order
(LO) of which turns out to start at $n=-3$, and ${\bf j}^{(n)}$ with LO
at $n=-2$.  These contributions are illustrated diagrammatically
for ${\bf j}^{(n)}$ in Fig.~\ref{fig:f1}.  The LO term involves the convection
and spin magnetization currents of individual nucleons (their
magnetic moments are taken from experiment), while the NLO one,
which scales $Q^{-1}$, is due to one-pion exchange (OPE) and
is well known---it is included in conventional calculations based on
meson-exchange phenomenology. At N2LO a relativistic correction
proportional to $1/m^2$---$m$ is the nucleon mass---to the
LO single-nucleon current occurs, while two-pion exchange (TPE) loop corrections begin at N3LO or
at order $Q$.  At N3LO there are also tree-level contributions
involving $\gamma\pi N$ vertices from sub-leading chiral Lagrangians,
as well as contact terms from minimal and non-minimal couplings
to the external EM field~\cite{Pastore09,Piarulli13}.

\begin{figure}[bth]
\begin{center}
\includegraphics[height=2.5cm]{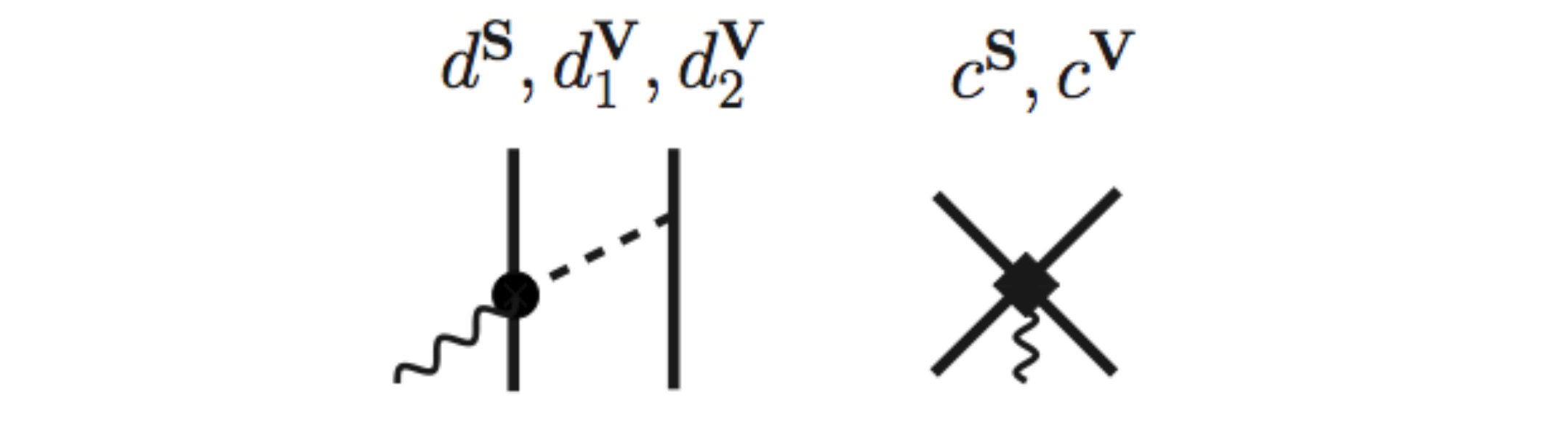}
\end{center}
\caption{The isoscalar $d^S$ and $c^S$, and isovector $d_1^V$, $d_2^V$,
and $c^V$ LEC's characterizing the current at N3LO.}
\label{fig:f2}
\end{figure}

There are 5 unknown LEC's that enter ${\bf j}$, see Fig.~\ref{fig:f2},
but none in $\rho$~\cite{Pastore09,Piarulli13,Koelling09,Koelling11}.  The operators derived so far
have power law behavior for large momenta, and need to be regularized
before they can sandwiched between nuclear wave functions.
The regulator is taken of the form $C_\Lambda (k)={\rm exp}[-(k/ \Lambda)^n]$
with $n=4$ and $\Lambda$ in the range 500-600 MeV.  Then the two isoscalar
LEC's are fixed by reproducing the deuteron and isoscalar trinucleon
magnetic moments, while two of the isovector LEC's, $d_1^V$ and
$d_2^V$ in Fig.~\ref{fig:f2}, are constrained by assuming $\Delta$-resonance
saturation~\cite{Piarulli13}.  Two different strategies have been adopted to fix the
remaining isovector LEC: it is determined by reproducing either the
$np$ radiative capture cross section $\sigma_{np}$ at thermal neutron energies
or the isovector trinucleon magnetic moment $\mu^V$~\cite{Piarulli13}.  There are no
three-body currents entering at the order of interest~\cite{Girlanda10}, and so it is
possible to use three-nucleon observables to fix some of these LEC's.
Their values are listed in Table~\ref{tb:t1}.  They are generally rather large,
particularly when $c^V$ is fixed by the $np$ radiative capture cross section.
The exception is the isoscalar LEC $d^S$ multiplying the one-pion exchange
current involving a sub-leading $\gamma \pi N$ vertex from the chiral
Lagrangian ${\cal L}^{(3)}_{\pi N}$.
\begin{table}
\begin{center}
\begin{tabular}{c||c|c||c|c|}
$\Lambda$  & $c^S$  & $d^S\times 10$ & $c^V(\sigma_{np})$ &$c^V(\mu^V)$ \\
\hline
500 &  4.1 & 2.2 &--13  &  --8.0\\
600  & 11 & 3.2 & --22&--12 \\
\hline
\end{tabular}
\end{center}
\caption{Values for the LEC's in units $1/\Lambda^2$
for $d^S$ and $1/\Lambda^4$ for $c^S$ and $c^V$; see text for further explanations.}
\label{tb:t1}
\end{table}

\section{Applications}
Next, we discuss applications of nuclear $\chi$EFT to the
few-nucleon systems, and defer to Saori Pastore's talk~\cite{Pastore15} for
results in the s- and p-shell nuclei.
The few-nucleon results are obtained from chiral two-nucleon
potentials developed by Entem and Machleidt at high order
in the power counting ($Q^4$ or N3LO)~\cite{Ent03,Machleidt11}, and chiral three-nucleon
potentials at leading order~\cite{Epelbaum02}.
Following a suggestion by Gardestig and Phillips~\cite{Gar06},
the two LEC's (in standard notation by now) $c_D$ and $c_E$
entering the latter have been constrained by fitting the
Gamow-Teller matrix element in tritium $\beta$-decay and the
binding energies of the trinucleons~\cite{Gazit09,Marcucci12}.
 
\subsection{Electromagnetic form factors of $A=2$--4 nuclei}
\begin{figure}[bth]
\includegraphics[width=15cm]{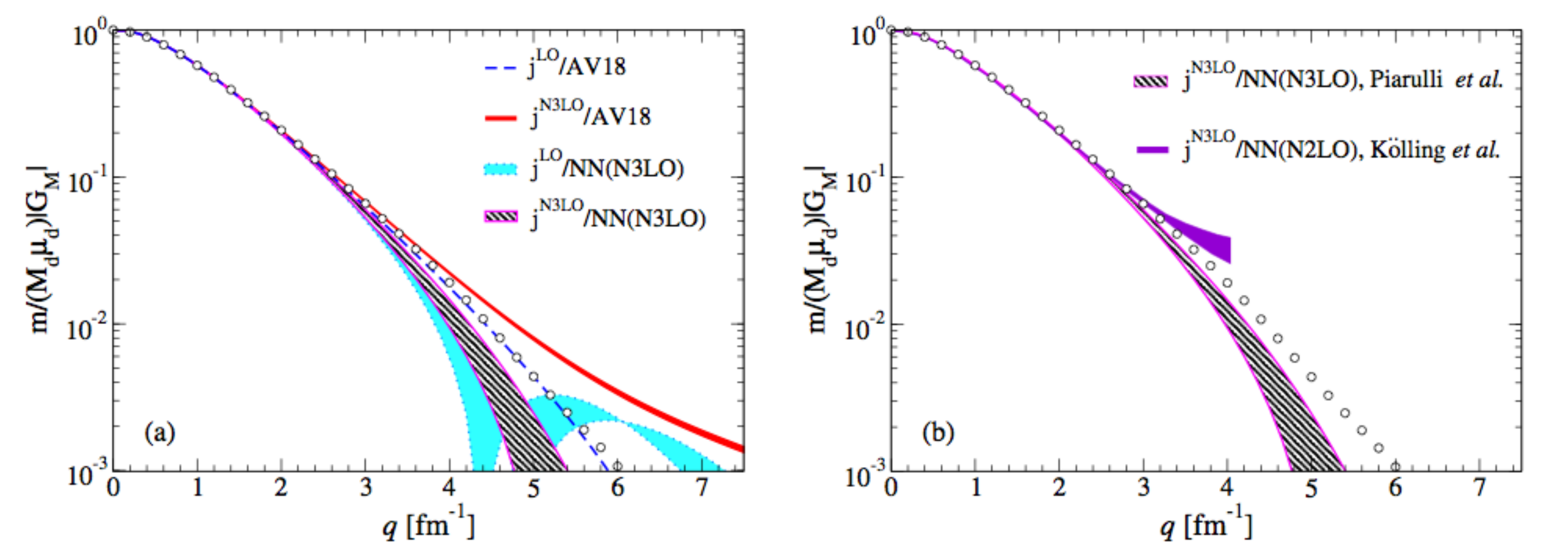}
\caption{Magnetic form factor of the deuteron: the left
panel shows results obtained with LO and N3LO currents and
either the chiral N3LO or conventional AV18 potential; the right panel shows
results obtained with N3LO currents and either the chiral N3LO
(same as in left panel) or a chiral N2LO potential by K\"olling {\it et al.}.
The bands reflect cutoff variation.  Experimental data are the empty
circles.}
\label{fig:f3}
\end{figure}

The deuteron magnetic form factor is shown in Fig.~\ref{fig:f3}~\cite{Piarulli13}.
The bands reflect the sensitivity to cutoff variations in the
range $\Lambda=500$--600 MeV.  The black bands include
all corrections up to N3LO in the (isoscalar) EM
current.  The NLO OPE and
N3LO TPE currents are isovector and therefore
give no contributions to this observable.  The right panel of Fig.~\ref{fig:f3} contains a
comparison of our results~\cite{Piarulli13} with the results of a calculation
based on a lower order potential and in which a different strategy
was adopted for constraining the LEC's $d^S$ and $c^S$
in the N3LO EM current~\cite{Koelling12}.
This figure and the following Fig.~\ref{fig:f5} are from the
recent review paper by S.\ Bacca and S.\ Pastore~\cite{Bacca14}.

\begin{figure}[bth]
\includegraphics[width=15cm]{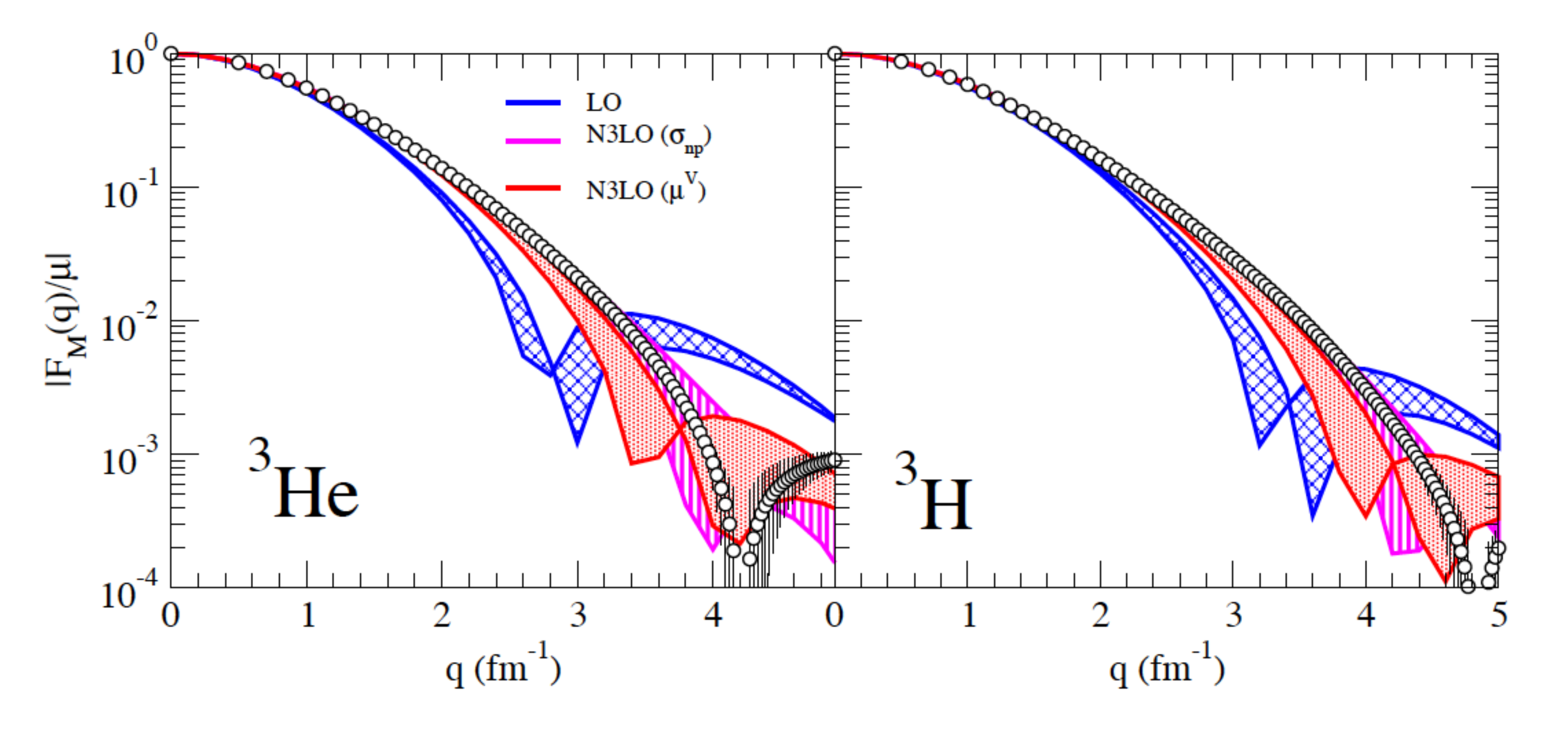}
\caption{Magnetic form factors of $^3$He (left panel) and $^3$H (right
panel); see text for further explanations.}
\label{fig:f4}
\end{figure}

The predicted magnetic form factors of the $^3$He and $^3$H ground states
are compared to experimental data in Fig.~\ref{fig:f4}~\cite{Piarulli13}.  Isovector two-body
terms in the EM current OPE and TPE play an
important role in these observables, confirming previous results obtained in the
conventional meson-exchange framework.  We show the N3LO
results corresponding to the two different ways used to
constrain the LEC $c^V$ in the isovector contact current
(recall the the LEC's $d_1^V$ and $d_2^V$ are assumed to be saturated
by the $\Delta$ resonance), namely by reproducing (i) the empirical value
for the $np$ cross section---curve labeled N3LO($\sigma_{np}$)---or
(ii) the isovector magnetic moment of $^3$He/$^3$H---curve labeled
N3LO($\mu^V$).  The bands display the cutoff sensitivity ($\Lambda=500$--600 MeV), which
becomes rather large for momentum transfers $q \gtrsim 3$ fm$^{-1}$.
The N3LO($\sigma_{np}$) results are in better agreement with
the data at higher momentum  transfers; however, they overestimate
$\mu^V$ by $\sim 2$\%.  On the other hand,
the N3LO($\mu^V$) results, while reproducing $\mu^V$ by construction,
underpredict $\sigma_{np}$ by $\sim 1$\% . 

\begin{figure}[bth]
\begin{center}
\includegraphics[height=7cm,width=9.5cm]{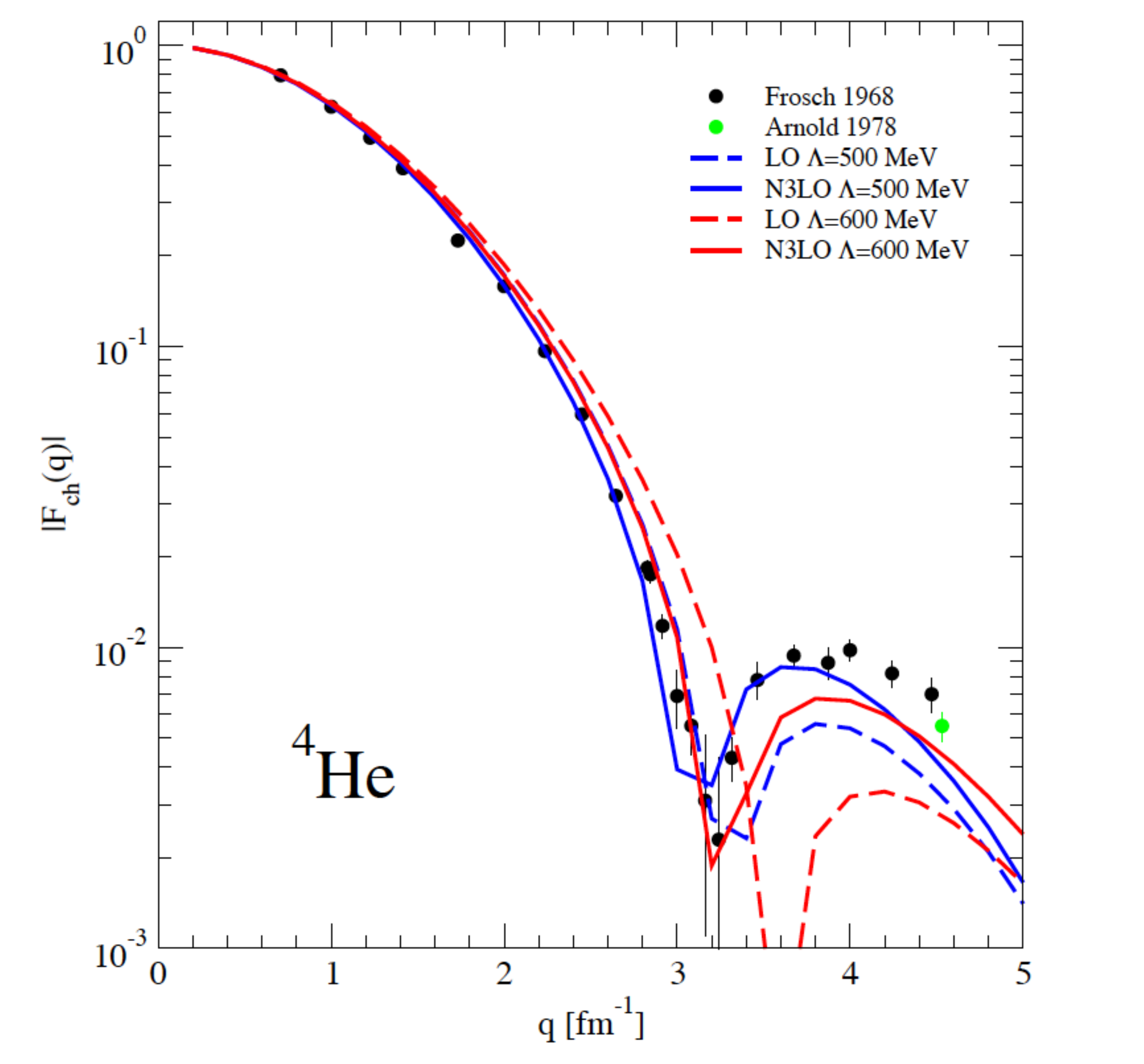}
\caption{The $^4$He charge form factor obtained from elastic
electron scattering data is compared to results obtained with the
LO and N3LO charge operator.}
\label{fig:f6}
\end{center}
\end{figure}

Moving on to the EM charge operator, we show in Figs.~\ref{fig:f5} and~\ref{fig:f6}
very recent calculations of the deuteron monopole and quadrupole form factors~\cite{Piarulli13} and
$^4$He (charge) form factor~\cite{Marcucci15}.  There are no unknown LEC's beyond $g_A$, $f_\pi$
and the nucleon magnetic moments---the latter enter a relativistic correction,
suppressed by $Q^2$ relative to the LO charge operator, {\it i.e.}, the well-known
spin-orbit term.  The loop contributions (at N4LO) from two-pion exchange are
isovector and hence vanish for these observables.
\begin{figure}[bth]
\begin{center}
\includegraphics[height=5.25cm]{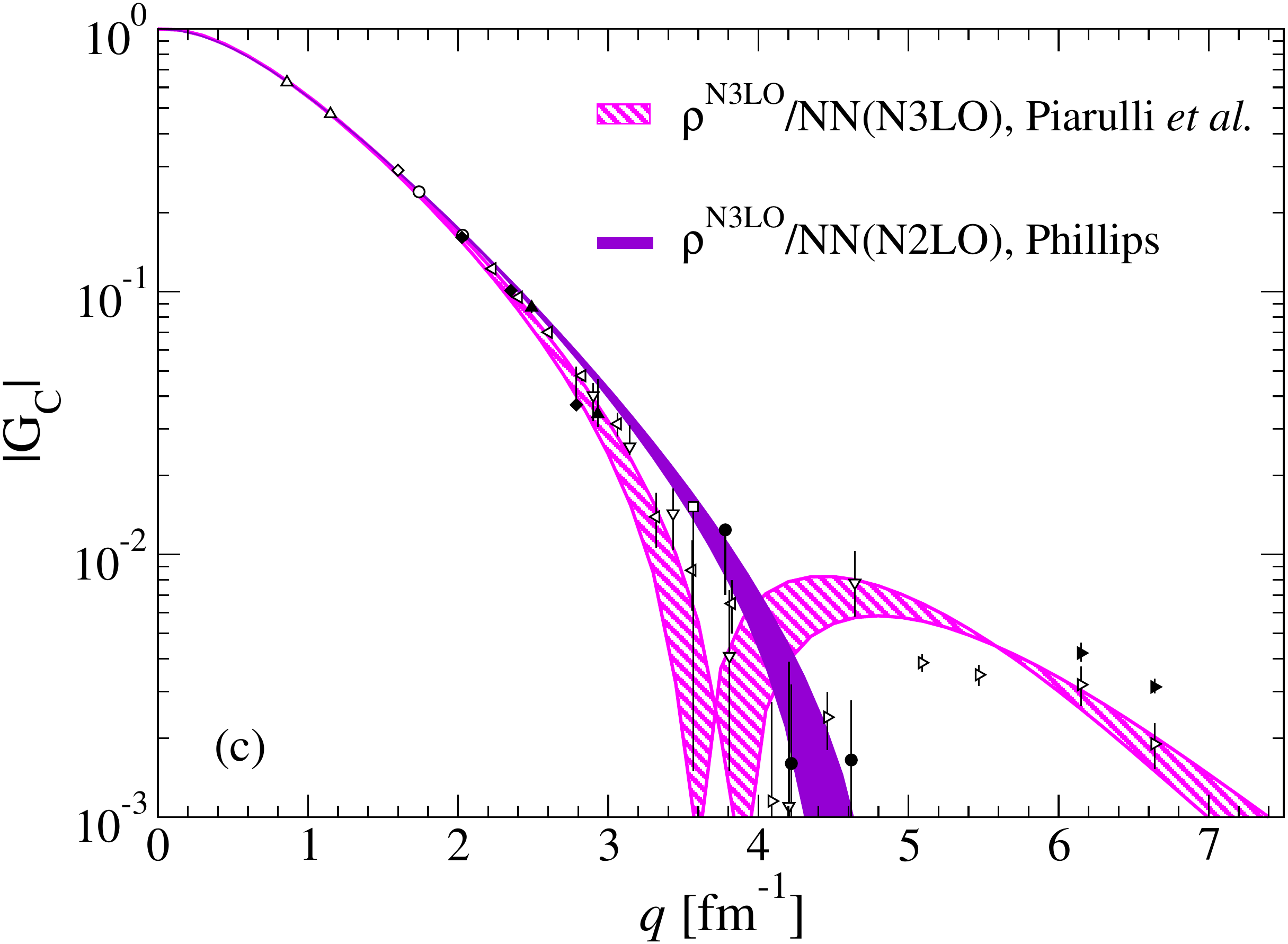}
\hspace{0.5cm}
\includegraphics[height=5.25cm]{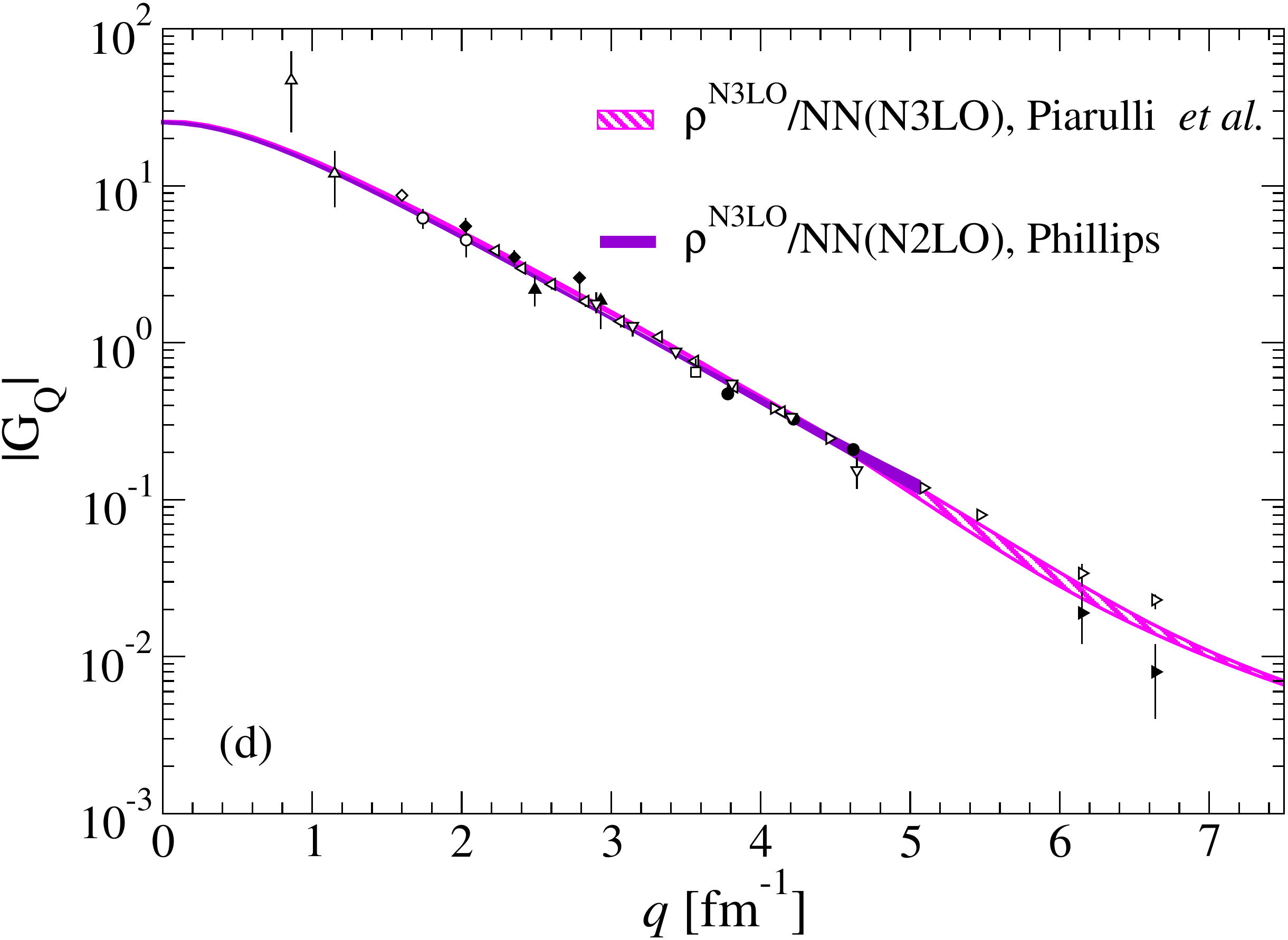}
\caption{The deuteron monopole and quadrupole form factors
obtained from measurements of the $A$ structure function and tensor polarization
are compared to predictions based on N2LO and N3LO chiral potentials.}
\label{fig:f5}
\end{center}  
\end{figure}

The deuteron monopole and quadrupole form factor data are obtained
from measurements of the $A$ structure function and tensor polarization
observable in electron-deuteron scattering.  In Fig.~\ref{fig:f5} the
two bands correspond to two different calculations, one of which,
labeled as NN(N2LO), is based on a lower order chiral potential~\cite{Phillips03,Phillips07}.
There is good agreement between theory and experiment.
Differences between the two sets of theory predictions merely
reflect differences in the deuteron wave functions obtained with the
N3LO and N2LO potentials. These differences are amplified in the diffraction region
of the monopole form factor.

The $^4$He charge form factor is obtained from elastic electron scattering
cross section data.  These data now extend up to momentum transfers
$q \lesssim 10$ fm$^{-1}$~\cite{JLABdata}, well beyond the range of applicability of
$\chi$EFT.  In Fig.~\ref{fig:f6} only data up to $q\lesssim 5$ fm$^{-1}$
are shown.  They are in excellent agreement with theory.

Predictions for the charge radii of the deuteron and helium isotopes and for the
deuteron quadrupole moment ($Q_d$) are listed in Table~\ref{tb:t2}~\cite{Piarulli13}.  They are within
1\% of experimental values.  It is worth noting that until recently calculations
based on the conventional meson-exchange framework used to consistently
underestimate $Q_d$. However, this situation has now changed, and a
relativistic calculation in the covariant spectator theory based on a one-boson
exchange model of the nucleon-nucleon interaction has led to a value for
the quadrupole moment~\cite{Gross15} which is in agreement with experiment.
\begin{table}
\begin{center}
\begin{tabular}{c|c|c|c|c}
      &  $r_c$($^2$H)  &  $Q_d$    & $r_c$($^3$He) & $r_c$($^4$He)  \\
      \hline
      & (fm)                 &     (fm$^2$) & (fm)                &   (fm)    \\
\hline
$\chi$EFT           &  2.126(4)        & 0.2836(16)    &  1.962(4)  &  1.663(11)      \\
\hline
EXP & 2.130(10)  & 0.2859(6) & 1.973(14)  & 1.681(4) \\
\hline
\end{tabular}
\end{center}
\caption{The charge radii of the $^2$H, $^3$He, and $^4$He nuclei,
and $^2$H quadrupole moment.}
\label{tb:t2}
\end{table}
\subsection{Weak transitions in few-nucleon systems}
Alessandro Baroni has discussed in this conference a recent derivation of the nuclear axial charge
and current operators up to one loop (N4LO) in the TOPT formalism
of Sec.~\ref{sec:chieft}~\cite{Baroni15}.  However, the results presented below have been
obtained at N3LO in these weak transition operators (no loops).
They are characterized by a single LEC in the axial current, which
we fix by reproducing the Gamow-Teller (GT) matrix element in tritium $\beta$-decay~\cite{Marcucci12}. 

A recent application of these transition operators is the calculation
of the rates for $\mu^{-}$ capture on deuteron and $^3$He~\cite{Marcucci12}.  These rates
have been predicted with $\sim 1$\% accuracy,
\[
\Gamma(^2{\rm H})=(399 \pm 3)\, {\rm sec}^{-1} \ , \qquad \Gamma(^3{\rm He})=(1494 \pm 21)\, {\rm sec}^{-1} \ .
\]
At this level of precision, it is necessary to also account for electroweak radiative corrections,
which have been evaluated for these processes in Ref.~\cite{Cza07}.  The error quoted on
the predictions above results from a combination of (i) the experimental error
on the $^3$H GT matrix element used to fix the LEC in the axial current,
 (ii) uncertainties in the electroweak radiative corrections---overall, these corrections
 increase the rates by 3\%--and (iii) the cutoff dependence.
 
There is a very accurate and precise measurement of the rate
on $^3$He: $\Gamma^{\rm EXP}(^3{\rm He})=(1496\pm 4)$ sec$^{-1}$~\cite{Ack98}.
We can use this measurement to constrain the induced pseudo-scalar
form factor of the nucleon.  We find $G_{PS}(q_0^2=-0.95\, m_\mu^2)=8.2 \pm 0.7$,
which should be compared to a direct measurement on hydrogen at PSI,
$G^{\rm EXP}_{PS}(q_0^2=-0.88\, m_\mu^2)=8.06 \pm 0.55$~\cite{Mucap},
and a chiral perturbation theory prediction of $7.99 \pm 0.20$~\cite{GPth}.

The situation for $\mu^-$ capture on $^2$H remains, to this day, somewhat confused:
there is a number of measurements that have been carried out,  but
they all have rather large error bars.  However, this unsatisfactory state of affairs
should be cleared by an upcoming measurement by the MuSun collaboration at PSI
with a projected 1\% error.

Another recent example is the proton weak capture on protons~\cite{Park03,Marcucci13}.
This process is important in solar physics: it is the largest source of
energy and neutrinos in the Sun.  The astrophysical $S$-factor
for this weak fusion reaction is one of the inputs in the standard
model of solar (and stellar) evolution~\cite{Bahcall04}.  A recent calculation based
on N3LO chiral potentials including a full treatment of
EM interactions up to order $\alpha^2$ ($\alpha$ is
the fine structure constant), shows that
it is now predicted with an accuracy of much less than 1\%:
$S(0)=(4.030 \pm 0.006) \times 10^{-23}$ MeV-fm$^2$.
\begin{figure}[bth]
\begin{center}
\includegraphics[height=7.5cm,width=12cm]{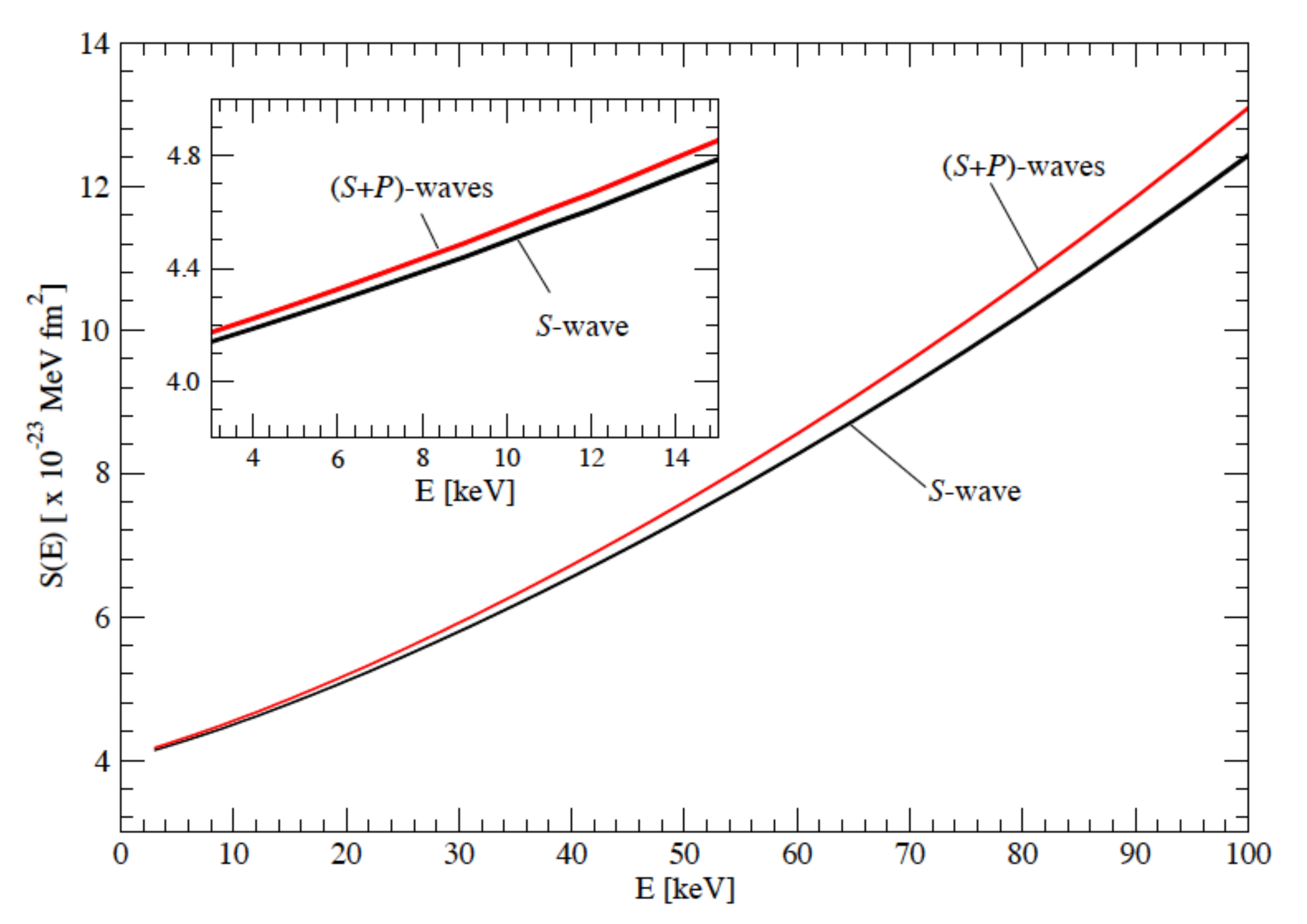}
\caption{The $S$-factor for $pp$ weak fusion due to S- and (S+P)-wave
capture as function of energy.}
\label{fig:f7}
\end{center}
\end{figure}
This calculation also included the (small) effects from
capture of the two protons in relative P-wave, see Fig.~\ref{fig:f7}~\cite{Marcucci13}.
The increase due to P-wave capture offsets the decrease
from higher order EM effects, in particular vacuum polarization.
\subsection{Hadronic weak interactions in few-nucleon systems}
Parity-violating (PV), but time-reversal invariant, chiral Lagrangians were
constructed by Kaplan and Savage~\cite{KS93} in the early nineties up to order $Q$,
\[
{\cal L}_{\rm PV}=
\overbrace{\frac{h^1_\pi}{2\sqrt{2}}\, f_\pi\, \overline{\psi}\, X^3_-\,\psi}^{{\cal L}_{\rm PV}^{(0)}} 
+{\cal L}^{(1)}_{\rm PV}\ ,
\]
where $h^1_\pi$ is the PV $\pi NN$ coupling.  More recently, their analysis
has been extended to order $Q^2$, and a complete listing of
isoscalar, isovector, and isotensor interactions up this order
has been provided in Ref.~\cite{Viv14}.  These Lagrangians imply
a PV potential that at order $Q$ depends on $h_\pi^1$ and 5 LEC's
(denoted as $C_i$ with $i=1,\dots, 5$) multiplying
short-range contact terms~\cite{Viv14,Gir08,Zhu05}.

In principle one needs 6 independent measurements to constrain
these LEC's.  A number of experiments has been performed:
measurements of the longitudinal asymmetry in
$\vec{p}$-$p$ scattering, of the photon asymmetry
in the $np$ radiative capture, and of the neutron spin rotation
in $\vec{n}$-$\alpha$ scattering.  Some are in progress, such
the measurement of the longitudinal asymmetry in the charge-exchange
reaction $^3$He($\vec{n}$,$p$)$^3$H, and some could be carried
out, such as the measurement of the neutron spin rotation
in $\vec{n}$-$d$ scattering.  These are beautiful, but difficult
experiments.  For example, in a neutron spin rotation experiment
one measures the rotation by an angle $\phi$ of the neutron spin in a plane perpendicular
to the beam direction~\cite{Schiavilla04},
\[
\frac{{\rm d}\phi}{{\rm d} d} = -\frac{2\pi\, \rho}{p}\, {\rm Re} \left[
M_+(p;\theta=0)-M_-(p;\theta=0)\right] \ .
\]
The effect depends on the density $\rho$ and length $d$ of the target, and on the
difference between the helicity + and -- forward scattering amplitudes ($p$ is the
relative momentum).  Its magnitude scales as $G_F \, f_\pi^2 \sim 10^{-7}$.

\begin{figure}[bth]
\begin{center}
\includegraphics[width=7.5cm]{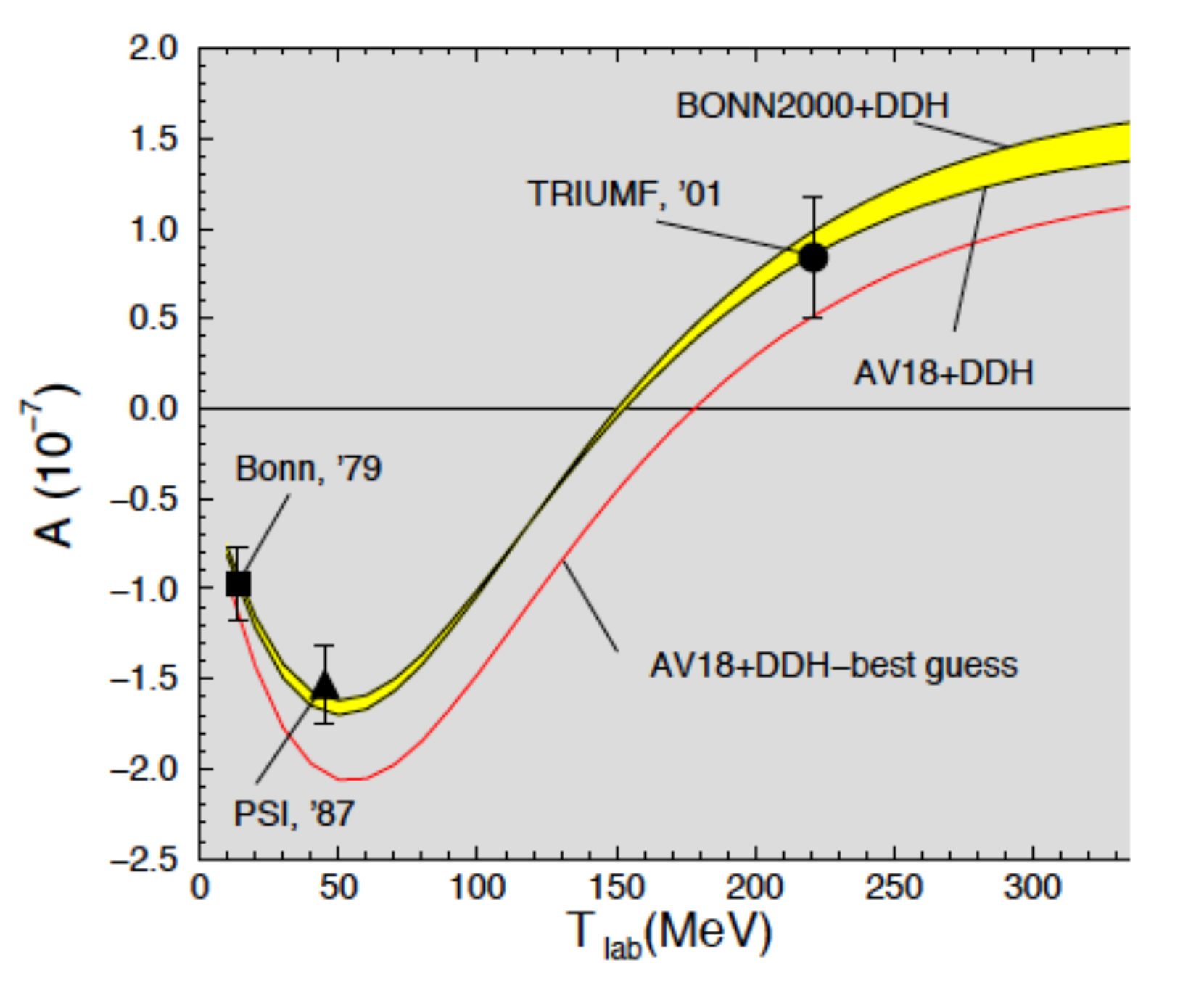}
\hspace{0.5cm}
\includegraphics[width=6.7cm]{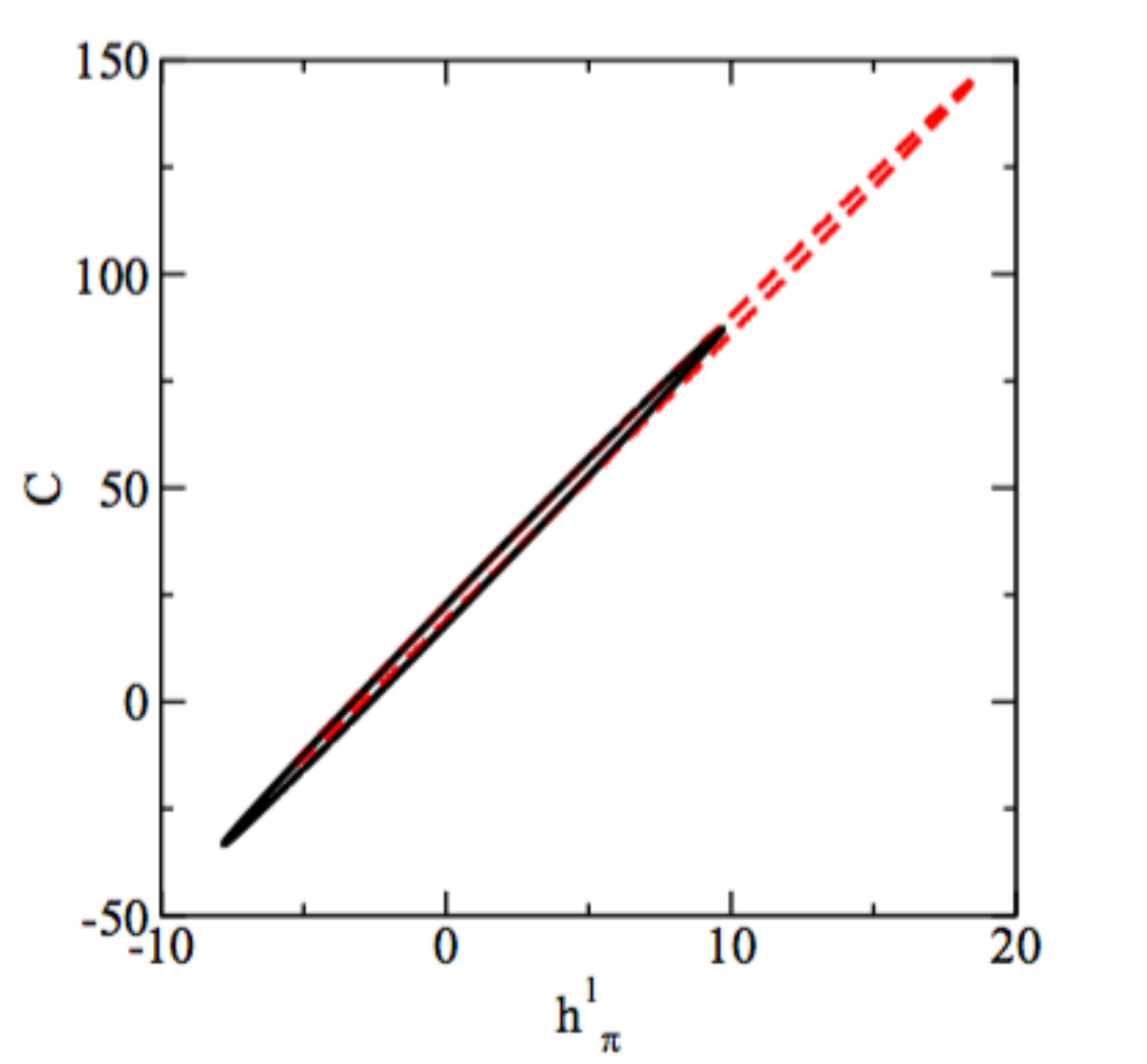}
\end{center}
\caption{The left panel shows the data compared to results
based on conventional strong-interaction potentials and the
DDH meson-exchange model for the parity-violating potential.
The right panel shows contour plots for the LEC's $h_\pi^1$ and $C$,
see text for further explanations.}
\label{fig:f8}
\end{figure}
\vspace{0.5cm}
In the following, we first discuss a recent attempt to constrain
$h_\pi^1$ by considering the longitudinal asymmetry in $\vec{p}$-$p$
scattering~\cite{Viv14}.  There are 3 data points, two at low energy and
one at relatively higher energy, see Fig.~\ref{fig:f8}.
Long-range contributions proportional to $h_\pi^1$ enter
via two-pion exchange, and the asymmetry can be written
as
\[
  A_z^{\, pp}(E,\Lambda) = a^{(pp)}_0(E,\Lambda) \, h^1_\pi + a^{(pp)}_1(E,\Lambda)\, C \ ,
\]
with $C$ denoting the linear combination of contact
LEC's $C= C_1+C_2 + 2\, (C_4+C_5)$.
The 3 experimental data points do not uniquely determine
$h_\pi^1$ and $C$: the right panel in Fig.~\ref{fig:f8}
shows contour plots  with a $\chi^2$-fit to these points.
One finds two very narrow and partially overlapping
ellipses corresponding to the cutoffs $\Lambda=500$ MeV
and 600 MeV.  There is a strong correlation between
$h_\pi^1$ and $C$, and a rather large range of variability
is allowed for these couplings.  It tuns out that the
two lowest data points are not independent,
in the sense that their energy dependence is driven
by that of the $^1$S$_0$ strong-interaction phase~\cite{Schiavilla04}.

\begin{table}[bth]
\begin{center}
\begin{tabular}{lcccccc}
\hline
$\Lambda$(MeV) &  $a_0$ & $a_1$ & $a_2$ & $a_3$ & $a_4$ & $a_5$ \\
\hline 
500  &   $-0.1444$  & $ 0.0061$ &  $ 0.0226$ &
         $-0.0199$  & $-0.0174$ &  $-0.0005$   \\
600  &   $-0.1293$  & $ 0.0081$ &  $ 0.0320$ &
         $-0.0161$  & $-0.0156$ &  $-0.0001$   \\
\hline
\end{tabular}
\end{center}
\caption{Values for the coefficients $a_i$ entering the parameter $a_z$.}
\label{tb:t3}
\end{table}
The next example is a recent analysis of the
(longitudinal) asymmetry in the charge-exchange reaction
$^3$He($\vec{n}$,$p$)$^3$H~\cite{Viv14}.  It is given by
 $a_z \, \cos\theta$, where $\theta$ is scattering angle
 and the parameter $a_z$ can be expressed as
 \[
  a_z  = a_0 \,h^1_\pi + a_1\, C_1 + a_2\, C_2
   + a_3\, C_3 + a_4\, C_4 + a_5\,  C_5 \ .
\]
The $a_i$ values are listed in Table~\ref{tb:t3}: $a_0$
is about an order of magnitude larger than $a_2$, the largest
among the coefficients multiplying the contact LEC's.
However, the (isoscalar) LEC $C_2$ is also expected to be
large.  Indeed, by matching the $C_i$ to the DDH estimates
for the PV vector-meson couplings~\cite{DDH}, one finds
\[
C^{({\rm DDH})}_1 \sim 1  \ ,\quad
C^{({\rm DDH})}_2 \sim  30  \ ,\quad
C^{({\rm DDH})}_3 \sim -2  \ ,\quad
C^{({\rm DDH})}_4 \sim 0  \ ,\quad
C^{({\rm DDH})}_5 \sim 7 \ .
\]
This indicates that the asymmetry $a_z\cos\theta$
results from the delicate balance between long-range OPE 
contributions and short-range contact contributions proportional
to $C_2$.
\section{Summary and outlook}

We have provided an overview of $\chi$EFT results for the
electroweak structure of few-nucleon systems.  There is
excellent quantitative agreement between experiment and
theory, at least in the region of low energy and momentum
transfer, where the $\chi$EFT approach is expected to be valid.
In some instances, such as in muon capture, results with 1\% accuracy
are obtained.  To echo the theme
of one of the talks~\cite{Epelbaum15} in this conference, we are entering
the precision era of nuclear $\chi$EFT.

Heavier nuclei offer new challenges and opportunities
for applications of nuclear
$\chi$EFT.  The quantum Monte Carlo methods
favored by our group to study nuclei with mass number $A >4$
are formulated in configuration space and, in particular,
configuration-space potentials are needed.  A first step in
this direction is the very recent development of a class
of minimally non-local two-nucleon chiral potentials
that fit the $np$ and $pp$ scattering database
up to lab energies of 300 MeV with $\chi^2$/datum
close to 1.3~\cite{Piarulli15}.

\vspace{0.5cm}

I wish to thank my collaborators A.Baroni, L.Girlanda, A.Kievsky, L.E.Marcucci,
S.Pastore, M.Piarulli, and M.Viviani for their many contributions to the
work presented in this talk.
The support of the U.S.Department of Energy under contract
DE-AC05-06OR23177 is also gratefully acknowledged.
\end{document}